\documentclass[10pt,preprint]{aastex}

\usepackage[]{graphicx}
\slugcomment{
To appear in ApJ Letters}

\begin{document}

\title{Discovery of a Near-Infrared Jet-Like Feature
in the Z Canis Majoris System}

\author{R. Millan-Gabet\altaffilmark{1} and J. D. Monnier\altaffilmark{2}}
\affil{Harvard-Smithsonian Center for Astrophysics, Cambridge, MA
02138}
\email{rmillan@cfa.harvard.edu,jmonnier@cfa.harvard.edu}

\altaffiltext{1}{now at: Caltech, Interferometry Science Center,
Pasadena, CA 91125}

\altaffiltext{2}{now at: University of Michigan, Ann Arbor, MI
48109}

\begin{abstract}

We present near-infrared high resolution observations of the young
binary system Z~Canis~Majoris using the adaptive optics system at the
Keck-II telescope. Both components are unresolved at 1.25\,$\mu$m and
1.65\,$\mu$m, although high dynamic range images reveal a previously
unknown jet-like feature in the circumstellar environment.  We argue
that this feature probably arises from light scattered off the walls
of a jet-blown cavity, and proper motion studies of this feature can
probe the dynamics of the bipolar outflow.  Potentially, the
morphology of the dust-laden cavity walls offers a new probe of the
momentum profile and collimation of bipolar winds from young stellar
objects.  We also derive high precision binary parameters, which when
combined with historical data have allowed the first detection of
orbital motion.  Lastly, our observations confirm the high degree of
flux variability in the system; the North-West binary component is
dominant at H-band, in contrast to all previous observations.

\end{abstract}


\section{Introduction}

Z~Canis~Majoris (Z~CMa) is a luminous and irregularly variable young
stellar object (YSO) originally classified as a Herbig~Ae/Be star
(HAeBe, intermediate to high mass pre-main sequence stars) on the
basis of its emission line spectrum and association with reflection
nebulosity \citep{Herbig:60}.  It underwent an outburst lasting a few
months in 1987, during which its spectral characteristics drastically
changed and its visual brightness increased by a modest $\sim0.7$
magnitudes \citep{Hessman:91}. High resolution optical and
near-infrared (IR) spectra in the post-outburst state showed features
typical of the FU~Orionis (FUOri) class, believed to be disk objects
undergoing a period of very active accretion
\citep{Hartmann:89}. Under this interpretation, Z~CMa is unique in
that it is the YSO with the highest known accretion rate ($\sim
10^{-3} M_{\sun} \cdot yr^{-1}$). The ranges of visual and near-IR
magnitudes of the unresolved system found in the literature are:
V=$11.20-8.80$, J=$6.16-5.87$, H=$5.09-4.70$ and K=$3.4-4.2$.

A companion was discovered in IR speckle observations by
\citet{Koresko:91}. The binary system had an angular separation of
$0.1\arcsec$ and position angle PA=$120\arcdeg$ (measured East from
North), where the primary is defined to be the North-West (NW)
component, which dominates the flux beyond $2.2\,\mu m$. Individual
spectral energy distributions reconstructed from the spatially
resolved photometry, combined with unresolved photometry at optical
and far-IR wavelengths, revealed that the emission in both components
is dominated by circumstellar material, i.e. there are no signs of
stellar photospheres.  The South-East (SE) component has been
identified as the FU~Ori object inferred by \citet{Hartmann:89}, while
the NW component is most-likely a HAeBe star surrounded by an
asymmetric dust envelope \citep[e.g.,][]{Garcia:99}.

In the last few years, HAeBe and FU~Ori objects have also been
resolved using the techniques of long baseline interferometry and
aperture masking at near-IR wavelengths \citep{Malbet:98,Akeson:00,
RMG:01,Tuthill:01,Danchi:01}. However, the interpretation of the
interferometer data depends on important model assumptions, namely the
extent to which the measured visibilities are systematically reduced
(which results in an overestimate of the size measured) by flux
arising in a widely-separated companion or scattered by distant dust.

The discovery reported here took place in the larger context of a
program of adaptive optics (AO) observations of HAeBe systems, aimed
at providing these constraints by exploring the level of extended
emission present at spatial scales intermediate between the very
narrow interferometric beams and the large scale nebulosity known to
surround these systems.

\section{Observations}

We observed Z~CMa on 2001~January~11, using the AO system
at the Keck-II telescope \citep{wiz:00}. These observations were
carried out using the slit viewing camera (SCAM) of the NIRSPEC
instrument \citep{mclean:00} as the imaging science detector.  In
these observations, Z~CMa itself -- unresolved by the Shack-Hartmann
sensor sub-apertures -- was used as the source of visible photons for
the AO wavefront sensor.  In order to achieve highest sensitivity to
scattered emission from circumstellar material, we limited our
observations to the shorter wavelength near-IR bands: J (N3 filter,
$\lambda=1.143-1.375 \, \mu m$) and H (N5 filter, $\lambda=1.413-1.808
\, \mu m$).

Our observing procedure consisted of establishing the maximum
integration time per frame which did not saturate the bright stellar
core, and co-adding 30 such frames. For all frames the correlated
double sampling (CDS) readout mode was used.  In order to search for
circumstellar emission beyond the normal field of view of the SCAM
array in AO mode ($4.3 \times 4.3 \arcsec$) and to improve the image
quality -- by removing the effect of the NIRSPEC slit -- we recorded 4
such frames nodding the telescope around a square by offsets of $4.3
\arcsec/2$, which resulted in a total field of view in the combined
image of $6.45 \times 6.45 \arcsec$. We also recorded J--band frames
in which the central core was allowed to saturate the detector, in
order to further augment our sensitivity to faint extended emission.

The results presented here have been obtained from the AO-corrected
images, after straightforward processing (i.e. no deconvolution has
been applied) consisting of: bad pixel removal, flat field correction
(using images of near-sunrise sky) and sky background subtraction
(using images obtained with a telescope offset of $60\arcsec$ from the
target source).

As an indication of the seeing quality and performance of the AO
system at the time of our Z~CMa observations, the images of an
unresolved calibrator star -- HD~54335, chosen to also have similar
visual and near-IR magnitude as Z~CMa -- were fitted with the sum of
two Gaussians of different widths, representing the narrow core and
wide halo into which partially corrected point spread functions (PSF)
can be approximately decomposed. Using the full-width at half-maximum
of the core component as a measure of effective seeing, we find on
average 45~mas and 46~mas at J and H bands respectively, both close to
the diffration limit of the 10~m telescope (31 and 42 mas
respectively). For this analysis, as well as in the rest of the paper,
stellar images are fitted with sub-pixel resolution using
2-dimensional pixelized Gaussians.

\section{The Z~CMa Binary}

We first derive the basic parameters of the Z~CMa binary from our
images. A model consisting of two scaled and displaced Gaussians gives
us the binary separation, PA, and flux ratio. These parameters are
derived for each of the N frames in our dithering procedure, and the
final value adopted is the mean of these independent measurements,
with a statistical error given by their standard deviation.

The plate scale and orientation of the AO images were calibrated using
observations of the quadruple system $\xi$ Ursae Majoris (HD~98230)
and its published orbital elements \citep{Mason:95}. Given that the
published orbit applies to the ``Aa'' center of mass and ``B''
component (the ``a'' component was unresolved until recent,
unpublished, Keck Aperture Masking observations in June 2000 by one of
the authors -- J. D. Monnier), we estimated the location of the former
in our images using the published masses \citep{Heintz:67}. Assigning
10\% and 30\% mass uncertainties to the ``A'' and ``a'' components
respectively, we obtain a plate scale of $17.32 \pm 0.06$~mas/pixel
(1.9\% different than nominal value of 17 mas/pixel) and a PA offset
(calculated $-$ nominal) of $-1.18 \pm 0.18 \arcdeg$.

These results are summarized in Table~\ref{bintable}.  In addition,
our 1.25\,$\mu$m and 1.65\,$\mu$m measurements are consistent with
both binary components being unresolved; that is, we find no evidence
for the extended component reported by \citet{Malbet:93} at somewhat
longer wavelengths ($\lambda>3.87\mu$m).

Figure~\ref{binfig2} shows the history of measurements of the Z~CMa
orbital elements up to this work
\citep{Koresko:91,Haas:93,Barth:94,Thiebaut:95} spanning a total of
11.2 years, and linear fits to the data.  While only a marginal change
in the angular separation is detected ($\chi^2_{reduced} = 0.3$ for
the linear fit, compared to 0.8 for the weighted mean), the PA has
clearly increased over this period of time ($\chi^2_{reduced} = 0.6$
for the linear fit, compared to 5.8 for the weighted mean), making
this the first detection of orbital motion in this system.

The linear fit to the PA data gives a total change of $8.8 \pm 1.5
\arcdeg$. For a distance to Z~CMa of about 1150~pc \citep{Herbst:87},
the $\sim 0.1 \arcsec$ angular separation corresponds to a projected
linear separation of 115~AU. Assuming masses for the components of $1
M_{\sun}$ and $5 M_{\sun}$, typical of T~Tauri and HAeBe stars
respectively, the period of a face-on circular orbit would be $\sim
500$ years, and the expected rate of position angle change $\sim 0.7$
degrees/year. Therefore, our detection of a $8.8 \arcdeg$ change in
about a decade is indeed plausible.

Previous workers have found the near-IR flux ratios to be highly
variable: $[NW/SE]_J=0.09-0.3$, $[NW/SE]_H=0.25-0.5$ and
$[NW/SE]_K=1.6-3.3$ \citep{Koresko:91,Haas:93}.  As can be seen in
Table~\ref{bintable}, we find the unexpected result that the NW
(HAeBE) component becomes brighter than the SE (FU~Ori) component at
H-band, in contrast to previous reports that the NW component became
the brightest in the system at K-band and longer wavelengths.  This
flux reversal could be naturally explained if the FU~Ori object has
been continuously fading following its recent outburst. Approximate
absolute photometry may be derived from our observations as follows.

The total flux of the Z~CMa system at our epoch is estimated using the
calibrator (HD~54335) as a flux standard. Its 2MASS fluxes are:
$F_J(\mbox{HD~54335}) = 19 \pm 4$~Jy and $F_H(\mbox{HD~54335}) = 31
\pm 6$~Jy. From our images we derive a ratio of fluxes between Z~CMa
and HD~54335 of $1.39 \pm 0.08$ and $2.68 \pm 0.25$ at J and H bands
respectively. The resulting combined fluxes of Z~CMa as well as those
of its components, derived using our estimates of the sum and ratio of
fluxes, are also given in Table~\ref{bintable}. Indeed, these
estimates clearly show that the FU~Ori object has been dimming
between 1986 and 2001. In particular, comparing with the epoch of the
\citet{Koresko:91} observations, the SE fluxes have decreased by $\sim
1.1 \pm 0.4$ mag and $\sim 0.7 \pm 0.3$ mag at J and H bands
respectively. In contrast, the HAeBe photometry is consistent with no
flux change at J band ($\Delta m_J \simeq 0.1 \pm 0.5$) and a
marginally significant increase at H band ($\Delta m_H \simeq 0.5 \pm
0.4$).

\section{High Dynamic Range Imaging: Discovery of a New Jet-Like Feature}
\label{section_cavity}
\subsection{Image Processing}

We have detected a new jet-like feature in deep images at J--band in
which the stellar cores were saturated.  Unfortunately, there are many
imaging artifacts in such high-dynamic range images, and we will
briefly mention the dominant ones before discussing the ``jet-like''
feature discovered around Z~CMa.  In Figure~\ref{ao1} we show the
mosaiced images of Z~CMa and the unresolved reference star (HD~54335),
in which an azimuthally-averaged PSF core has been subtracted to
artifically enhance the dynamic range in the image.  We have marked
the dominant artifacts common to the both Z~CMa and the reference
star, but note the interesting new feature {\it not} associated with
the PSF.  Prior to this detection, no extended structure was known for
Z~CMa at these wavelengths and scales ($< 1$ arcsec) beyond that of a
pure binary \citep[nebulosity on larger scales has been reported
by][]{Nakajima:95}.

Figure~\ref{jetfig} shows a close-up of the region near the new narrow
feature, rotated so that North is up and East is left.  In order to
remove the wings of the PSF, we have subtracted a two-component model
of the binary star using the PSF measured with the reference star.
Clearly this subtraction is imperfect, as evidenced by the speckled
residuals within 0.5$\arcsec$ of the binary.  Despite these obvious
flaws, the dynamic range (DR) in the image (unsaturated peak to noise
ratio) is $\sim$ 200000 beyond $\sim 0.5 \arcsec$ from the central
binary, and the signal-to-noise in the jet-like feature is greater
than 10.  The jet-like feature extends to $\sim 0.9\arcsec$ (1135~AU)
from the central binary in the S-SW direction, and has a width of
$\sim 0.1\arcsec$ (115~AU).

\subsection{Discussion}
It is likely that the extended emission we have detected is stellar
light or disk emission scattered off of dust grains, since at the
observed distance from the central stars this dust would be too cold
to emit thermally in the near-IR. Alternatively, the emission could
arise from a faint emission line within J--band bandpass, a
possibility which needs to be eliminated in follow-up observations. We
note that the jet-like feature is not detected in our H--Band images,
likely a consequence of the lower DR in those images (with
non-saturated peaks) and/or the lower efficiency of scattering by dust
at longer wavelengths.

Assuming scattering by dust, an intriguing interpretation is that this
scattering takes place off the inner walls of an evacuated cavity
carved by the well-known large-scale outflow.  A bipolar outflow was
discovered by \citet{Poetzel:89}, traced by several Herbig-Haro
objects (blue and red-shifted sides) and a jet (blueshifted side, also
SW direction). The total extent of the outflow was 3.6 pc, oriented at
PA=$60\arcdeg$.  \citet{Garcia:99} reported an optical jet in [OI]
extending approximately 1$\arcsec$ from Z~CMa at PA $240\arcdeg$.  The
radio counterpart of the outflow was also detected in CO lines
\citep{Evans:94} and in the thermal jet \citep{Velazquez:01}, both
oriented coincident with the optical jet.  In the latter case, the
high angular resolution of the VLA observations allowed for the first
time to clearly identify the origin of the outflow with the SE
(FU~Ori) binary component.  Understanding the shape of the jet-blown
cavity and the amount of swept-up dust could constrain the momentum
profile of the bipolar winds, testing current theories for such winds
and the generation of jets \citep[e.g.,][]{Shang:98,Matzner:99}.

Figure~\ref{jetfig} also indicates the direction of the previously
known optical jet and bipolar outflow at PA=$240\arcdeg$.  If the
near-infrared feature arises from light scattering off dust swept-up
into a thin shell surrounding a jet-blown cavity, we would expect
features on either side of the jet direction (as is also indicated in
Figure~\ref{jetfig}) and both should be (more-or-less) static.
However, we do not see any evidence for symmetrical emission on the
other side of the optical jet, casting doubt on this hypothesis.  On
the other hand, the high level of spurious emission may be masking
this feature, and even higher dynamic-range observations are needed
before this interpretation can be ruled out.

Alternatively, the emission may be directly associated with a
different or new jet, given its morphology and association with
known accretion objects.  If the feature is moving with typical jet
velocities (e.g., 500 km/s, Poetzel~et~al.~1989), then proper motions
of the jet feature of up to 100~AU per year ($\sim 90$ mas per year)
are expected and should be readily detectable within a short
time-span, in contrast to the previous hypothesis.

Unfortunately, due to the PSF artifacts in these first images, the new
feature cannot be followed close enough to the central binary to
establish whether or not it is associated with one of the components,
and higher signal-to-noise at close separations plus optimum placement
of the feature with respect to the PSF artifacts may be able resolve
this issue.

\section{Conclusions}

We have detected a new jet-like feature in the close environment of
the young stellar binary Z~CMa. The feature extends about 1035~AU from
the central binary, in the S-SW direction, and has a width of about
115~AU. A symmetric feature to the other side of the known optical jet
and bipolar outflow is expected but not detected, and therefore
further targeted observations are required to establish whether this
new feature is indeed associated with the bipolar outflow
(e.g. scattering off dust in its inner walls) or constitutes an
independent jet. Our high resolution observations have also revealed
for the first time orbital motion in the ZCMa system, as well as
further established the high degree of flux variability of its
individual components.

\acknowledgments

R.M.G acknowledges that this work was performed while he was
a Michelson Postdoctoral Fellow, funded by the Jet Propulsion
Laboratory, which is managed for NASA by the California Institute of
Technology.  J.D.M. acknowledges support from a Center for
Astrophysics Fellowship at the Harvard-Smithsonian Center for
Astrophysics. The data presented here were obtained at the W. M. Keck
Observatory, which is operated as a scientific partnership among the
California Institute of Technology, the University of California, and
the National Aeronautics and Space Administration. The observatory was
made possible by the generous financial support of the W. M. Keck
Foundation. This publication makes use of data products from the Two
Micron All Sky Survey, which is a joint project of the University of
Massachusetts and the Infrared Processing and Analysis
Center/California Institute of Technology, funded by the National
Aeronautics and Space Administration and the National Science
Foundation.

\clearpage

\figcaption[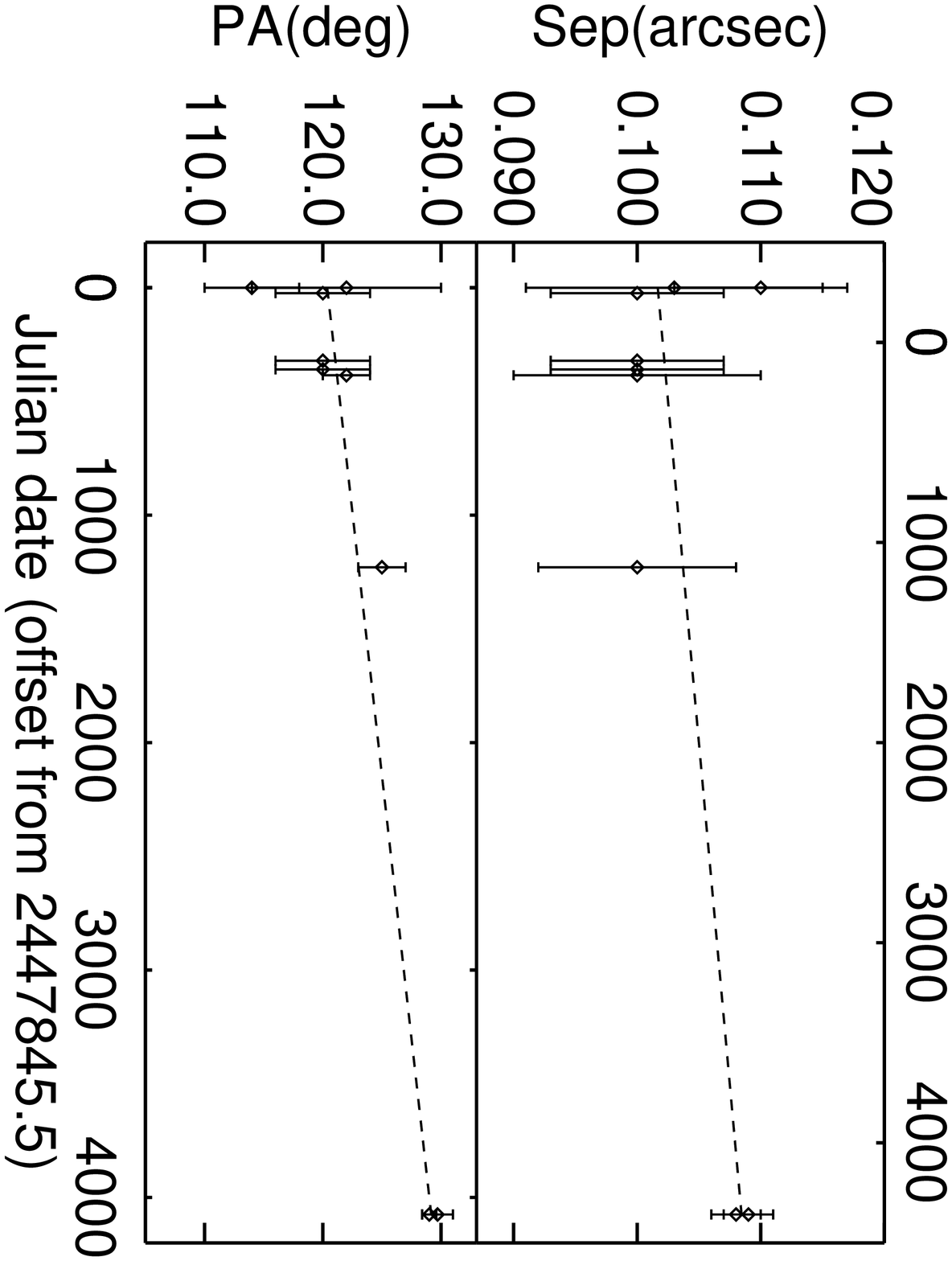]{Chronology of Z~CMa binary separation measurements.
We clearly detect a position angle increase of $8.8 \pm 1.5 \arcdeg$
over 11.2 years.
\label{binfig2}}

%% comment these out for submission
\begin{figure}[h]
\begin{center}
\includegraphics[angle=90,width=0.8\columnwidth]{f1.eps}
\end{center}
\end{figure}
\clearpage

\figcaption[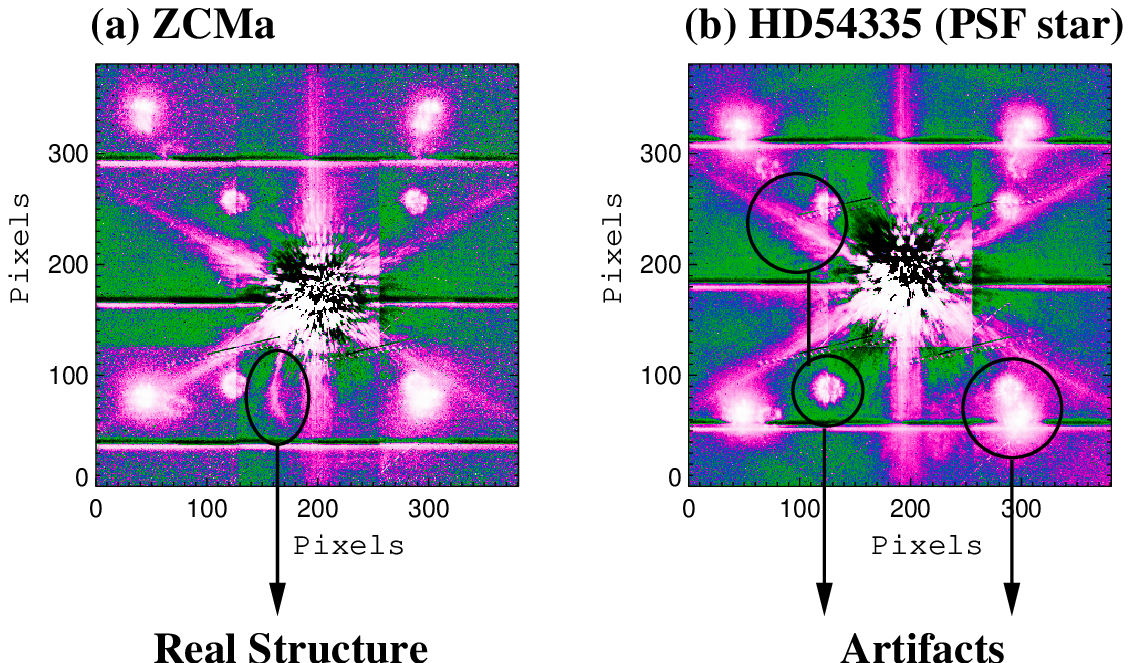]{High dynamic range images of Z~CMa at J-band.
Panels~(a) \& (b) show the mosaiced frames where the new jet is
clearly visible amid the PSF artifacts, which repeat in the calibrator
observation. The most prominent PSF artifacts are diffraction spikes
and internal camera reflections, repeated four times in these images
as a result of the mosaicing procedure. We have chosen to display
our images in a way that emphasizes the intrinsic detector and optical
artifacts, so that the fidelity of the
plume structure can be accurately judged.
\label{ao1}}

%% comment these out for submission
\begin{figure}[h]
 \begin{center}
\includegraphics[angle=0,width=\columnwidth]{f2.eps}
\end{center}
\end{figure}
\clearpage

\figcaption[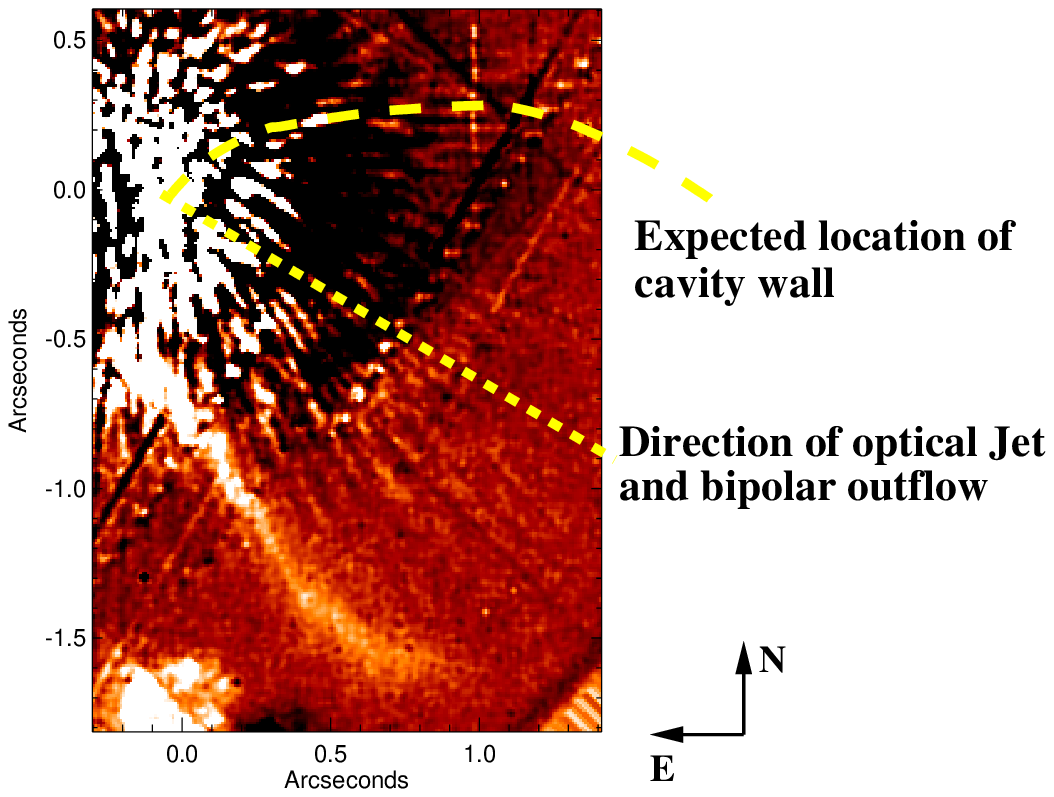]{High dynamic range images of Z~CMa at J-band,
centered on the the region near the new jet-like feature. The binary
components have been subtracted out of this image, and the color table
has been stretched to show the low surface brightness features of the
images. This figure also indicates the direction of the known optical
jet and bipolar outflow. If the jet-like feature is marking one
wall of the jet-blown cavity, we expect to find emission of the cavity
wall on the other side of the jet at the location indicated (see
discussion in \S\ref{section_cavity}).
\label{jetfig}}

%% comment these out for submission
\begin{figure}[h]
\includegraphics[angle=0,width=\columnwidth]{f3.eps}
\end{figure}

\clearpage

\begin{deluxetable}{cccccccc}
\tabletypesize{\scriptsize}
\tablecaption{Measurements of binary parameters and approximate photometry \label{bintable}}
\tablewidth{0pt}
\tablehead{
\colhead{Filter} & \colhead{N}   &
\colhead{Flux Ratio (NW/SE)} & \colhead{Separation ($\arcsec$)} & 
\colhead{PA ($\arcdeg$)} & Total Flux (Jy) & Flux SE (Jy) & Flux NW (Jy)}
\startdata
J(N3)		& 4 & 0.50 $\pm$ 0.05 & 0.109 $\pm$ 0.002 & 129.0 $\pm$ 0.6 & 2.7 $\pm$ 0.6  & 1.8 $\pm$ 0.4 & 0.9 $\pm$ 0.2\\
H(N5) 		& 3 & 1.18 $\pm$ 0.06 & 0.108 $\pm$ 0.002 & 129.7 $\pm$ 1.3 & 7.9 $\pm$ 1.7  & 3.6 $\pm$ 0.8 & 4.3 $\pm$ 0.9\\
\enddata
\end{deluxetable}

\end{document}